\documentstyle[sprocl,epsf]{article}

\def\NPB{{\em Nucl. Phys.} B}
\def\PLB{{\em Phys. Lett.}  B}

\def\PRD{{\em Phys. Rev.} D}
\def\ZPC{{\em Z. Phys.} C}

\def\be{\begin{equation}}
\def\ee{\end{equation}}
\def\bea{\begin{eqnarray}}
\def\eea{\end{eqnarray}}
\def\Vec#1{\mbox{\boldmath $#1$}}

\begin{document}

\title{ON THE COLOUR CONFINEMENT\\ AND THE MINIMAL SURFACE}

\author{Sadataka Furui }

\address{School of Science and Engineering, Teikyo University, 
Utsunomiya,\\ 320-8551, Japan\\E-mail: furui@dream.ics.teikyo-u.ac.jp} 

\author{Bilal Masud }
\address{Centre for High Energy Physics, Punjab University, Lahore, \\ 54590
 Pakistan\\E-mail: bilalmasud@usa.net}


\maketitle\abstracts{In the analysis of the energy of the four-quark system obtained in the SU(2)
lattice Monte Carlo, the f-model in which the transition potential is expressed
in the form $f=f_ce^{-k_Ab_sA-k_P\sqrt{b_s}P}$, where A is the area and P is
the perimeter of the Wilson loop, was successful in the 
case of simple configurations of the four quarks. In the case of tetrahedral 
geometry, an estimation of the minimal surface whose contours run the positions
of the four quarks is necessary. We show that the regular surface approximation
whose area can be calculated analytically, is a good approximation for 
evaluating the minimal surface. The numerical value of the coefficient $k_Ab_s$ is close to $2 fm^{-2}$ which is the density of
the ${\Vec Z}_2$ vortex in the SU(2) lattice Monte Carlo.}

One of the most significant manifestations of the colour confinement in
QCD is the area law. 
In 1994, Helsinki group measured the four-quark energies in quenched 
SU(2) lattice Monte Carlo at $\beta=2.4$ and $2.5$ on $16^3\times 32$ lattice
\cite{GMS}.
Quarks are infinitely heavy, but via rearrangement of the flux tubes, the
two two-quark systems interact with each other and  energy spectra of 
various kinds of four-quark configurations were measured. The energies were
analysed by a method similar to the resonating group method based on Hamiltonian QCD. In this theory, the
transition potential between different 4-quark configurations
is expressed by 
\begin{equation}
V=\left(\begin{array}{cc} v_{13}+v_{24} & (f/N_c)V_{AB}\\
                      (f/N_c)V_{AB} & v_{14}+v_{23}\end{array}
\right) 
\end{equation}
where 
\begin{equation}
f=f_ce^{-k_Ab_sA-k_P\sqrt{b_s}P}
\end{equation}
Here, $b_s$ is defined from the string tension, A is the area and
P is the perimeter of the minimal surface\cite{FGM,Pen}.  
The effective string action of QCD is expected to have the Nambu-Goto
area term and Jacobian that comes from the field variables to the string 
variables\cite{Poly}. The long-range dynamics is dominated by the area term, 
and we include corrections by the perimeter dependent term\cite{MM}. 

We consider minimal surface spanned by four fixed points that makes a 
tetrahedron. If the conformality is ignored, a surface defined by four
fixed points in ${\bf R}^3$ can be achieved
 by taking the surface spanned by straight lines ${\Vec X}_u$
and ${\Vec X}_v$, where ${\Vec X}(u,v)$ is defined as 
\begin{equation}
{\Vec X}(u,v)=u(1-v){\Vec r}_{41}+v(1-u){\Vec r}_{42}+uv{\Vec r}_{43}\quad
(regular 2)\nonumber
\end{equation}
 or
\begin{equation}
{\Vec X}(u,v)=u(1-v){\Vec r}_{14}+v(1-u){\Vec r}_{34}+uv{\Vec r}_{24}\quad
(regular 1)\nonumber
\end{equation}
where the ${\Vec r}_{ij}={\Vec r}_i-{\Vec r}_j$ is the vector between position
of quarks. Here $12$ are on the bottom and $34$ are on the top, $13$ are on the left and $24$ are on the right.

The surface spanned by straight lines is called the regular surface\cite{DoC},
and its area is given as
\begin{equation}
area=\int_0^1 du\int_0^1 dv |{\Vec X}_u\times {\Vec X}_v|
\end{equation}
which can be expressed as an analytical function of parameters $r$ and $d$. 

Since the regular surface is not the minimal surface, 
we performed variation such that the mean curvature becomes as small as
possible. The ansatz that we adopt is,
\begin{equation}
{\Vec X}'(u,v)={\Vec X}-t m(u,v){\Vec k}
\end{equation}
\begin{equation}
m(u,v)=uv(1-u)(1-v)[1+c uv(1-u)(1-v)]H_n(u,v)
\end{equation}
where
\begin{equation}
H_n(u,v)=-2 {\Vec X}_u\cdot {\Vec X}_v({\Vec X}_u\times {\Vec X}_v)\cdot {\Vec X}_{uv}
\end{equation}

\begin{figure}[htb]
\begin{minipage}[b]{0.47\linewidth} 
\begin{center}
\epsfysize=140pt\epsfbox{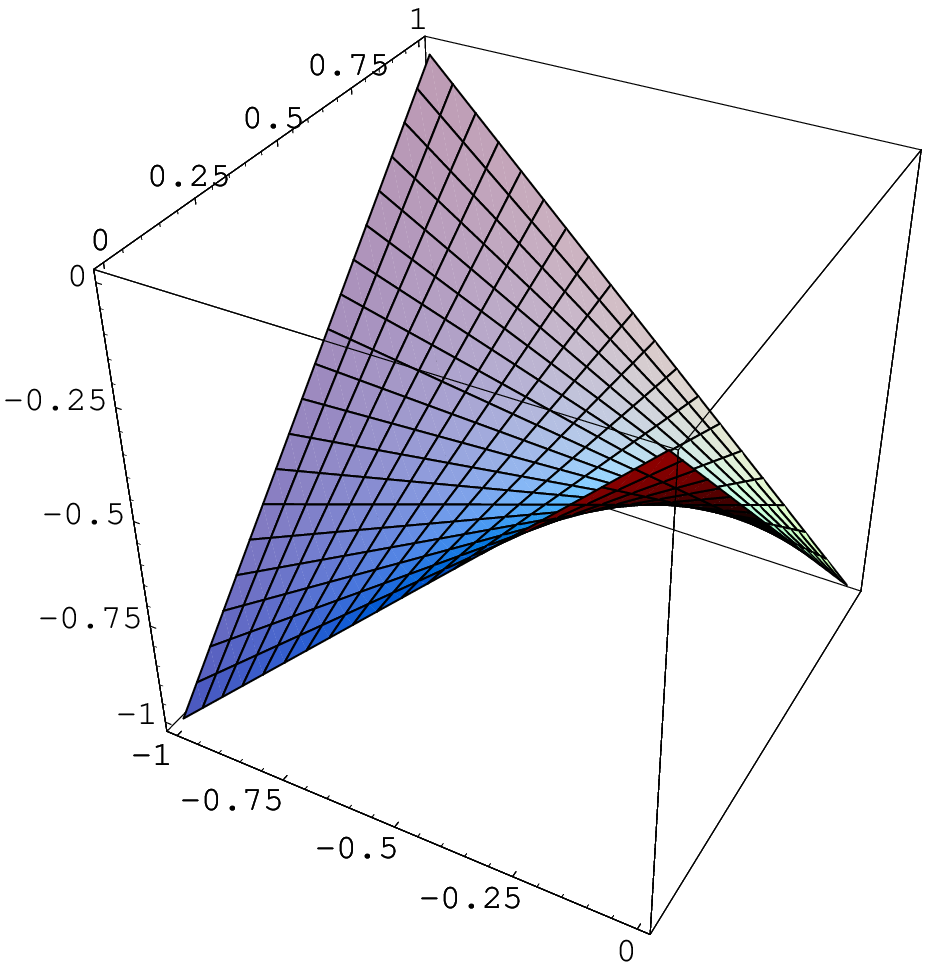}
\caption{The regular surface. The area is $1.2808$. $(r=d=1)$}
\end{center}
\end{minipage}
\hfil
\begin{minipage}[b]{0.47\linewidth} 
\begin{center}
\epsfysize=140pt\epsfbox{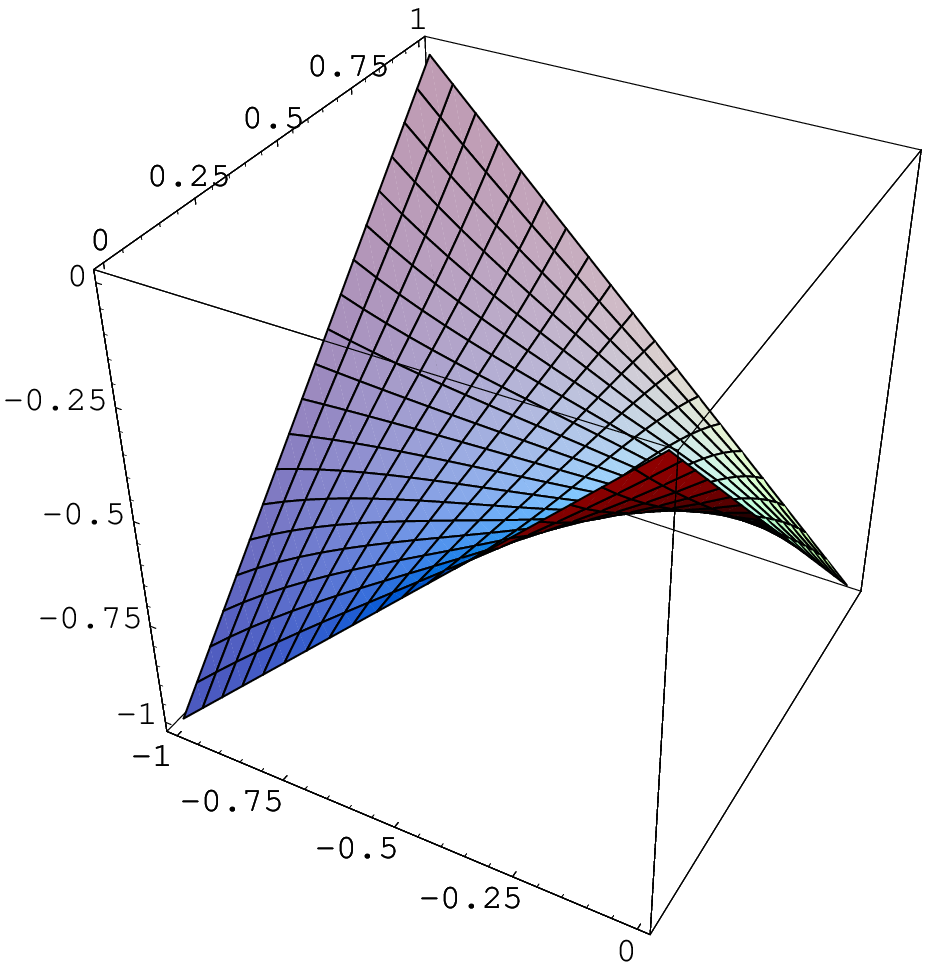}
\caption{The 'minimal' surface with four straight line boundary condition. The area is $1.2793$. $(r=d=1)$}
\end{center}
\end{minipage}
\end{figure}

\begin{table}[hb]
\begin{tabular*}{\textwidth}{@{}l@{\extracolsep{\fill}}cccc}

\hline
& c & t & area \\
\hline
 &0 & 0 & 1.2808 \\
 &0 & 0.48364 & 1.2794 \\
 &-5.97149 & 0.49226 & 1.2793 \\
\hline
\end{tabular*}
\caption  {The area calculated by variation. $(r=d=1)$}
\end{table}

As shown in the Table 1, and the Figure 1,2, the difference between the
regular surface and the variationally obtained non-regular surface is
small.  Examples of other terahedra were also calculated.

If one relaxes the boundary condition, the minimal surface for 4 fixed points
can be obtained by the conformal mapping. In the study of a string whose
ends run with the light velocity, a minimal surface whose 
pair of opposite boundaries are straight lines but the other pairs are free, 
which is called the 'Gergonne's surface', was studied\cite{Poly}.
The conformal mapping which yields the Gergonne surface of a tetrahedron is 
given by
$x=-u$, $y=\cos u \sinh v/(\cos(\pi/4)$ $\sinh(\pi/4))$, $z=\sin u \sinh v/(\sin(\pi/4) \sinh(\pi/4))$, whose 3d-plot is shown in Fig.3. Due to the curved bound
ary, the area of this surface $13.807$ is 
larger than that of the corresponding $regular2$ surface, which is $11.062$. The area of
the $regular1$ surface, which has different boundaries is $8.761$.

When one relaxes all the four boundaries, the conformal mapping in the lowest
non-trivial order 
 $x=u v/3$, $y=-v+(3u^2 v-v^3)/108$, $z=u+(u^3-3u v^2)/108$, yields the minimal
surface.  An example shown in Fig. 4 has the area $49.4$, which is larger than 
that of regular surface $36\times 1.2808=46.12$. 

In the effective string theory, the short distance dynamics is not defined by
the area term alone, but the difference of the area of $regular1$ and 
$regular2$ surface allows to judge which type of fluxtube breaking is 
preferred.
\begin{figure}[htb]
\begin{minipage}[b]{0.47\linewidth} 
\begin{center}
\epsfysize=120pt\epsfbox{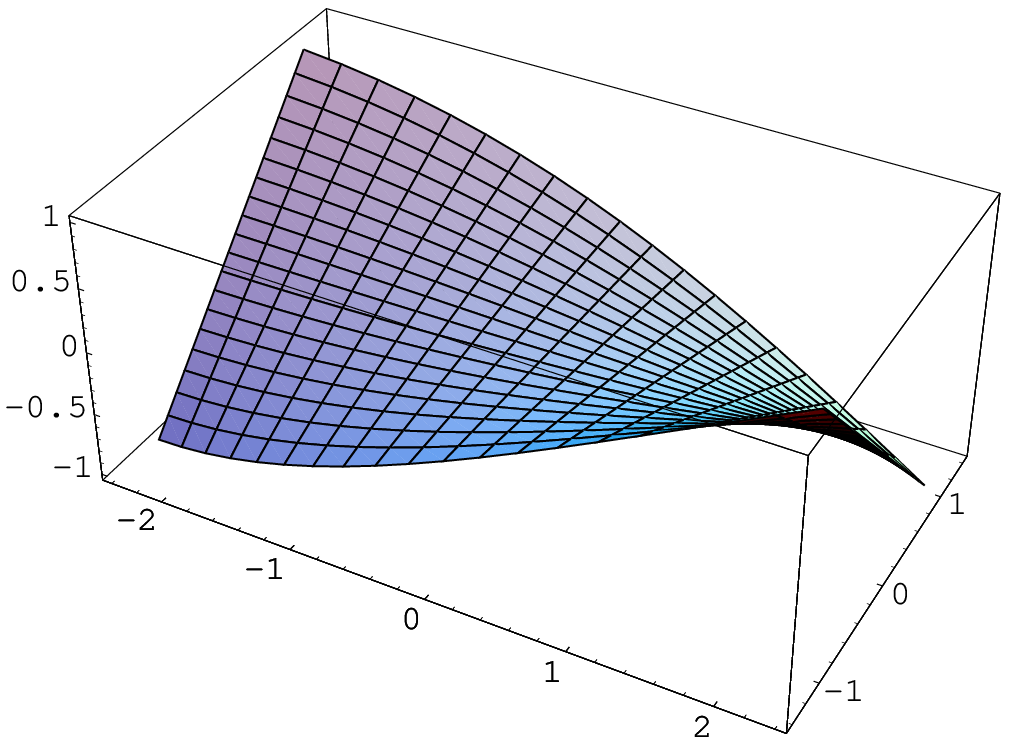}
\caption{The minimal surface with two straight line boundary condition. The area is $13.81$. $(r=3\pi/2,d=2)$}
\end{center}
\end{minipage}
\hfil
\begin{minipage}[b]{0.47\linewidth} 
\begin{center}
\epsfysize=120pt\epsfbox{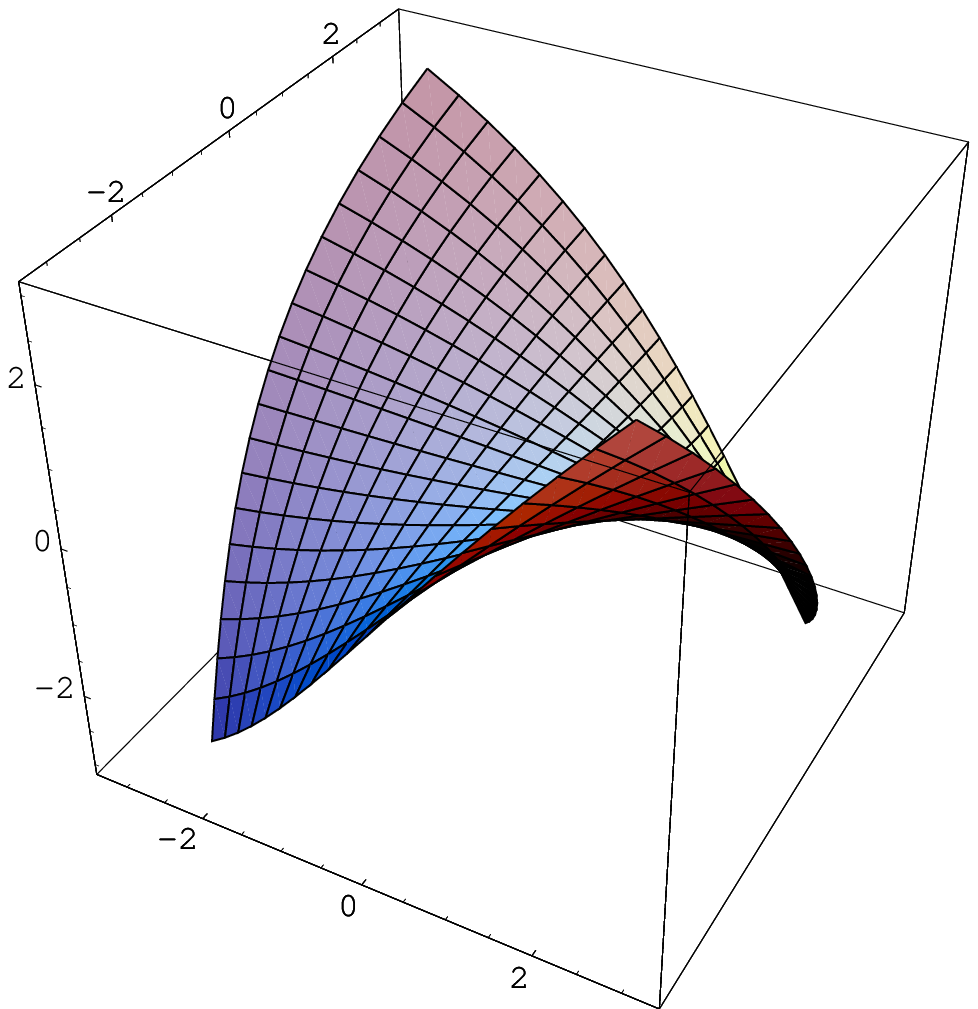}
\caption{The minimal surface with curved boundary condition. The area is $49.4$.
 $(r=d=6)$}
\end{center}
\end{minipage}
\end{figure}

\begin{table}[hb]
\begin{tabular*}{\textwidth}{@{}l@{\extracolsep{\fill}}cccccccc}
\hline

    & $\beta$ &$k_A$    & $b_s a^2$  & $a$(fm) & $k_A b_s(fm^{-2})$ &$k_P$ & \\
\hline
 &2.4 &0.296(11)  & 0.0724  &  0.12 & 1.5 & 0.080(2) & FGM \\
\hline
 &2.4 & 0.38(4)  & 0.0709  & 0.1194(9) & 1.89 & 0.087(10) &Pennanen  \\
 &2.5 & 0.73(8)  & 0.0373 & 0.0866(9) & 3.63 & 0.008(13) &Pennanen  \\
\hline
\end{tabular*}
\caption  {Parameters of the f-model.}
\end{table}

The f-model for the four-quark system of simple configurations indicates 
that the perimeter contribution reduces as $\beta$ becomes large\cite{Pen}.  
The parameters used in the f-model are summarized in
Table 2. The value of $k_A b_s$ is about $2 fm^{-2}$ which is close
to the density of the ${\Vec Z}_2$ centre vortex in the SU(2)\cite{Rht,step,KT}.
Since in the change of the topology of flux tubes, piercing of a thick centre 
vortex
is expected to play a role, this numerical coincidence is suggestive, but
we need further investigation on the role of the 
centre vortex in the confinement.

S.F. is grateful to Dr. Polykarpov for the information on the string theory and
the minimal surface.

\end{document}